# Improving Deep Learning Performance for Predicting Large-Scale Porous-Media Flow through Feature Coarsening

Bicheng Yan*, Dylan Robert Harp, Bailian Chen, Rajesh J. Pawar

Earth and Environmental Sciences, Los Alamos National Laboratory, Los Alamos, NM, USA, 87544

*Corresponding author

Email: bichengyan@lanl.gov (B. Yan); dharp@lanl.gov (D.R. Harp); bailianchen@lanl.gov (B. Chen); rajesh@lanl.gov (R.J. Pawar)

*Abstract* - Physics-based simulation for fluid flow in porous media is a computational technology to predict the temporal-spatial evolution of state variables (e.g. pressure) in porous media, and usually requires high computational expense due to its nonlinearity and the scale of the study domain. This letter describes a deep learning (DL) workflow to predict the pressure evolution as fluid flows in large-scale 3D heterogeneous porous media. In particular, we apply feature coarsening technique to extract the most representative information and perform the training and prediction of DL at the coarse scale, and further recover the resolution at the fine scale by 2D piecewise cubic interpolation. We validate the DL approach that is trained from physics-based simulation data to predict pressure field in a field-scale 3D geologic $CO_2$ storage reservoir. We evaluate the impact of feature coarsening on DL performance, and observe that the feature coarsening can not only decrease training time by >74% and reduce memory consumption by >75%, but also maintains temporal error <1.5%. Besides, the DL workflow provides predictive efficiency with ~1400 times speedup compared to physics-based simulation.

## I. INTRODUCTION

Fluid flow in porous media drives the overall performance of many geologic $CO_2$ storage (GCS) and energy extraction processes (Benson and Surles, 2006; Chen et al., 2006; Chen et al., 2018). In physics-based simulations, the governing equations used to describe fluid flow in porous media can be accurately discretized by traditional numerical methods (Chen et al., 2006) and thus the temporal-spatial evolution of state variables (e.g., pressure and saturation) in the porous media can be accurately predicted. The main drawback of this approach is the high computational expense, which is associated with the scale of the geological model and the nonlinearity due to heterogeneities (Yan et al, 2016), complex fluid thermodynamics (Michael et al., 2018), and coupled physics processes (Winterfeld and Wu, 2016; Zhan et al., 2020).

As researchers have demonstrated deep learning's (DL) superior capability to process high dimensional data (Georgious et al., 2020) and approximate various continuous functions (Csaji, 2001), many recent research studies focus on enhancing the capability to predict the evolution of state variables in porous media with DL. A family of physics-informed neural network (PINN) models impose physics governing equations to regularize the loss function during the training process through automatic differentiation (Raissi et al., 2019; Fuks and Tchelepi, 2020; Harp et al., 2021). These approaches ensure that the neural networks are consistent with the physics governing fluid flow in porous media. While PINN is suitable for predicting processes governed by physics with medium complexity, its computation may become expensive to solve problems with large scale and high nonlinearity. Alternatively, image-based approaches have also been investigated to predict fluid flow in porous media, and mainly leverage convolutional neural networks (CNN) to approximate the nonlinear relationship between geological maps (e.g., permeability) and flow maps (e.g., saturation) in fluid flow in porous media (Zhong et al., 2019; Tang et al., 2020). With sparse connectivity between input and output, image-based approaches tend to be more appropriate to deal with medium-scale heterogeneous porous media. However, for capturing heterogeneity with high resolution, the scale of a realistic geological model can easily reach millions of grid cells, which is extremely challenging for existing DL methodologies to digest the data for temporal and spatial regression tasks. So far there is little work to evaluate the training efficiency and predictive accuracy of the evolution of state variables in such large heterogeneous geological models.

Here, we describe a DL workflow to predict the evolution of pressure as fluid flows in large-scale 3D geological models. This workflow falls into the category of image-based approaches, as it takes full advantage of the spatial topology predictive capability (Hughes et al., 2018; Huang et al., 2020) of CNN, specifically Fourier Neural Operator (FNO) (Li et al., 2020). Our main contributions are twofold. Firstly, we apply feature coarsening technique to extract the most representative information from the input maps (e.g., permeability and porosity), and perform the training and prediction of FNO at the coarse scale. This can significantly decrease the memory consumption of the training data and computational cost for training, and thus makes DL more affordable for large scale geological models. Secondly, based on the hypothesis of pressure continuity, we further recover the resolution of predicted pressure fields to the original fine scale through 2D piecewise cubic interpolation approach. To validate the performance, we apply the DL workflow that is trained from physics-based simulation data to predict the pressure evolution in a field-scale 3D heterogeneous geological $CO_2$ storage reservoir, and provide comprehensive analysis to assess the memory efficiency, training efficiency and predictive accuracy of the DL workflow.

## II. PHYSICS CONTEXT

At a geologic $CO_2$ storage site where $CO_2$ is injected into a saline aquifer, the fluid phases include a water-rich (liquid) phase and a $CO_2$-rich (supercritical) phase, and the primary components in the water-rich and $CO_2$-rich phases are water and $CO_2$, respectively. The flow and transport of each component in the porous media is governed by the mass balance equation,

$$\frac{\partial}{\partial t}\left(\phi \sum_\alpha S_\alpha \rho_\alpha x_{\alpha,i}\right) - \nabla \cdot \left\{K \sum_\alpha \frac{k_{r\alpha}}{\mu_\alpha} \rho_\alpha x_{\alpha,i}(\nabla p_\alpha + \rho_\alpha g \nabla Z)\right\} + \sum_l \left(\sum_\alpha \rho_\alpha x_{\alpha,i} q_\alpha\right)^l = 0, \qquad (1)$$

where the first term is the fluid storage, the second is the advective flux based on Darcy's law, and the third is the source or sink term. Subscript $i$ indicates the fluid component, including water and $CO_2$; $\alpha$ indicates the fluid phase, including water-rich phase $w$ and $CO_2$-rich phase $g$; $t$ is time; $\phi$ is the rock porosity; $S_\alpha$ is the phase saturation; $\rho_\alpha$ is the fluid phase density; $x_{\alpha,i}$ is the mole fraction of component $i$ in fluid phase $\alpha$; $K$ is the permeability of porous media; $k_{r\alpha}$ is the phase relative permeability; $\mu_\alpha$ is the phase viscosity; $p_\alpha$ is the phase pressure; $g$ is the acceleration due to gravity; $Z$ is depth; $q_\alpha$ denotes the rate for extracting or injecting fluid phase $\alpha$ through well perforation $l$.

Additionally, Equation (1) is constrained by several auxiliary relationships, including the equality between fluid volumes and pore volume, capillary pressure constraint and fluid thermodynamics equilibrium. Additional details can be found in previous literature (Chen et al., 2006; Michael et al., 2018). In a physics-based simulator, the mass balance equation combined with the auxiliary relationships are solved iteratively to calculate the state variables.

## III. METHODOLOGY

The aim of the DL workflow is to predict pressures evolving globally with time in a 3D porous media, providing a computationally efficient alternative to physics-based simulators. In the section, we illustrate the details of the DL workflow including feature selection and assembly, feature coarsening, deep neural network architecture, training, prediction and resolution recovery.

**A. Feature Selection and Assembly**

Here, the features are the input variables to predict the evolution of pressure as fluid flows in porous media. To be consistent with the data structure of the 2D convolutional neural network architecture we apply in this work, the features are also assembled as 2D images.

In Equation (1), the permeability $K$ characterizes the degree of spatial connectivity of fluid flow. Since the vertical permeability $K_V$ is usually much lower than the horizontal permeability $K_H$, e.g., $\frac{K_V}{K_H} = 0.1$ in our work, the vertical connectivity contributes much less to the fluid flow than the horizontal connectivity. Therefore, we omit the impact of vertical connectivity and slice the 3D permeability volume into 2D horizontal layer-wise images. Additionally, the porosity $\phi$ only contributes to the fluid storage, so we can also slice 3D porosity volumes into 2D horizontal layer-wise images. The fluid phase rates $q$ are functions of time, and are also variables to control the fluid flow. Since there will only be a limited number of fluid injection or production wells, the 2D feature image of fluid phase rates is filled with zeros everywhere except the location where the wells are drilled. Finally, a feature image of time is included for temporal evolution of the state variable pressure ($p$), and to flexibly perform interpolation at arbitrary time steps.

As we consider strong permeability heterogeneity, e.g., the permeability ranges from $10^{-4}$ to $10^4$ millidarcy in our work, we logarithmically scale the permeability. Therefore, the features to predict the state variable pressure ($p$) includes logarithmic permeability $log(K)$, porosity $\phi$, fluid phase rate $q$ and time $t$.

**B. Feature Coarsening**

Assembly of the 2D feature images as a feature array will consume large quantities of memory, which can potentially lead to memory allocation issue and low training efficiency. Therefore, we plan to coarsen the image size to a certain level before proceeding to the training process. In this work, the feature images are coarsened by selecting spatial points with a constant stride, which is defined as an increment value added to the preceding spatial point in order to generate the next spatial point, while we always keep the most representative information, such as high permeability and porosity zones and injection well locations. Correspondingly, the output image of pressure is also processed in the same way.

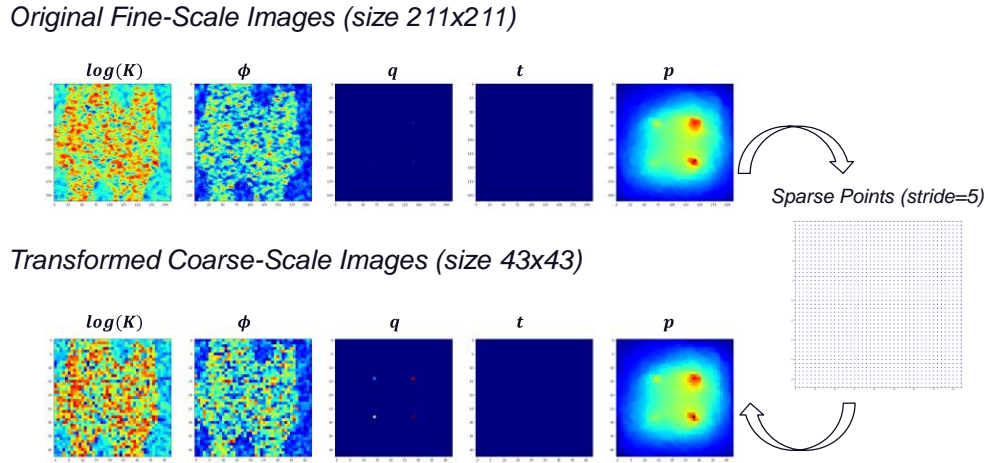

**Fig. 1. Transformed images of feature and state variables from fine-scale to coarse-scale.**

An example of coarsening process is depicted in **Fig. 1**. By taking a stride of 5, the fine-scale images of feature and state variables with size 211×211 are transformed into coarse-scale images with size 43×43. During this process, the most representative information is well preserved in the coarsened images. The high permeability and high porosity zones, the nonzero injection rate values at the well locations and the pressure plume are still captured at coarse scale.

**Table 1** presents that as we increase the stride from 1 to 10, the size of the 2D images decreases from 211×211 to 22×22, and the corresponding memory allocated to the feature array decreases by 2 orders of magnitude (from 53.50 gigabytes to 0.58 gigabytes). The sensitivity analysis clearly demonstrates that the feature coarsening strategy significantly saves memory consumption.

Table 1. Sensitivity of stride with memory consumed by the feature array (precision: float 32)

| Stride | Feature Image Size | Feature Array Dimension* | Consumed Memory (gigabytes) | Normalized Memory (%) |
|---|---|---|---|---|
| 1 | 211×211 | 80,640×211×211×4 | 53.50 | 100.00% |
| 2 | 106×106 | 80,640×106×106×4 | 13.50 | 25.24% |
| 3 | 73×73 | 80,640×73×73×4 | 6.40 | 11.97% |
| 4 | 55×55 | 80,640×55×55×4 | 3.64 | 6.79% |
| 5 | 43×43 | 80,640×43×43×4 | 2.22 | 4.15% |
| 10 | 22×22 | 80,640×22×22×4 | 0.58 | 1.09% |

*Feature array dimension is number of samples × image width × image length × number of feature types.

**C. Deep Neural Network (DNN) and Training Process**

As we tackle large-scale heterogeneous geological models, an image-based approach is preferred to regress the high dimensional problem. The DNN we adopted in this work is the Fourier Neural Operator (FNO) proposed by Li et al. (2020), which is demonstrated to have superior predictive capability for physics-based processes. In FNO, the feature array $X$ is first mapped into a high dimensional representation $V_0$ through a fully connected dense layer. Further, $V_l$ is recursively updated through,

$$V_l = \sigma\big(WV_{l-1} + \kappa(V_{l-1})\big), l = 1, \dots, L. \qquad (2)$$

where $V_l$ are the feature maps at layer $l$, and is a function of $V_{l-1}$ preceding it; $\sigma$ denotes the nonlinear activation function; $W$ is a linear operator defined by a 1D convolutional operator; $\kappa$ is a 2D convolution operator defined in the Fourier space. Ultimately, $V_L$ is transformed back to the state variable $p$ through several fully connected dense layers. For more theoretical context about FNO, refer to Li et al. (2020). The architecture of FNO is depicted in **Fig. 2**. To simplify the illustration, the layer related to the operation in Equation (2) is labeled as a "Fourier" layer, and the fully connected layer is named as a "FC" layer. FNO takes 4 input images ($log(K), \phi, q, t$) at layer FC-1, then sequentially goes through 4 Fourier and 2 FC layers to predict the output image of state variable $p$.

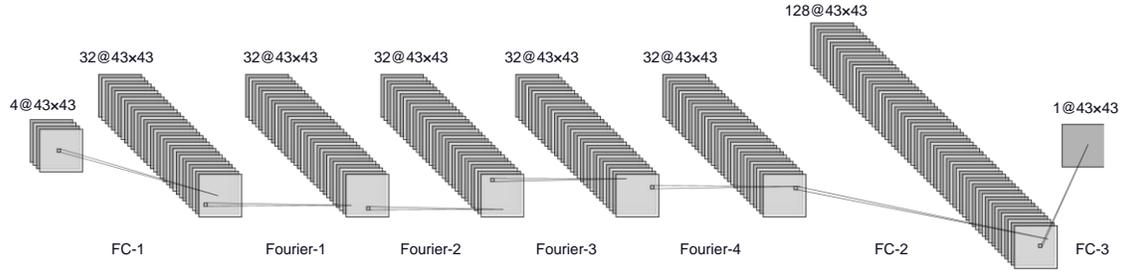

**Fig. 2. Architecture of Fourier Neural Operator with 3 fully connected (FC) and 4 Fourier layers. "$a@n_x \times n_y$" at the top of each layer denotes: $a$ - number of features; $n_x$ - the image width; $n_y$ - the image length.**

In **Fig. 2**, there is no activation function $\sigma$ involved in FC-1 and FC-3 layers, and the $\sigma$ used through Fourier-1 to Fourier-4 is a LeakyReLU,

$$\text{LeakyReLU} = \begin{cases} x & if\ x \geq 0 \\ 0.01 & otherwise \end{cases}, \tag{3}$$

$\sigma$ in FC-2 is ReLU,

$$\text{ReLU} = \begin{cases} x & if\ x \geq 0 \\ 0 & otherwise \end{cases}, \tag{4}$$

Different from other convolutional neural networks, there is no pooling layer in FNO for down sampling and summarizing the average or the most activated presence of the feature maps. The loss function $\mathcal{L}$ in our FNO approach is defined as,

$$\mathcal{L}(\theta) = \|p - \hat{p}\| + \lambda \|p_w - \hat{p}_w\|, \tag{5}$$

where $\theta$ are the learnable parameters; $\|\cdot\|$ is the root-mean-square-error operator; $p$ is the ground truth of the pressure; $\hat{p}$ is the prediction of $p$ by FNO; $\lambda$ is a weighting factor; $p_w$ is the ground truth of $p$ at the well locations; $\hat{p}_w$ is the prediction of $p_w$ by FNO. The second term in Equation (5) is a regularization term for enhancing the resolution at the well locations.

The goal of training FNO is to find $\theta$ by minimizing the loss function $\mathcal{L}$. We implemented FNO and the associated training module with the deep learning library PyTorch (Paszke et al., 2019), and adopted the Adam optimizer to train FNO.

**D. Recover the Resolution of Prediction at Fine Scale**

FNO performs prediction at the coarse scale, but in tasks like data assimilation (Jafarpour et al., 2010), we need to predict pressures at monitoring wells which are defined at the original fine scale but may not be available at the coarse scale. Hence, in such scenarios it necessitates to recover the resolution of FNO prediction in order to flexibly predict pressures at arbitrary locations in the domain.

The resolution recovery is based on spatial interpolation. During feature coarsening, we select the spatial points with a pre-defined stride, so the coarse-scale spatial coordinates ($x^c, y^c$) can be easily tracked based on the fine-scale spatial coordinates ($x^f, y^f$) and the stride, where superscripts $f$ and $c$ denotes fine and coarse scales, respectively. Assume that the pressure $p^f$ at the fine scale is spatially continuous, then we can recover $p^f$ by interpolation based on the predicted pressure $p^c$ by FNO at the coarse scale and the coordinates $(x^c, y^c)$ and $(x^f, y^f)$,

$$p^f = \mathcal{F}(p^c, x^c, y^c, x^f, y^f), \tag{6}$$

where $\mathcal{F}$ is an interpolation operator, and the interpolation scheme adopted by this work is the 2D piecewise cubic interpolation (Virtanen et al., 2020). The hypothesis of pressure continuity will hold in most cases, unless there is a no-flow boundary such as a sealing fault traversing the geological model, which induces local pressure discontinuity nearby the boundary. However, this type of scenario is not in the scope of this work.

## IV. NUMERICAL EXPERIMENTS

### A. Reservoir Model Description

We use a 3D heterogeneous reservoir model with 211 by 211 by 30 grid cells in the $x$, $y$ and $z$ directions respectively, and a total number of grid cells of 1,335,630. The grid dimension is 500 $ft$ by 500 $ft$ by 10 $ft$ in the $x$, $y$ and $z$ directions, respectively, and the grid cell size is uniform throughout the domain.

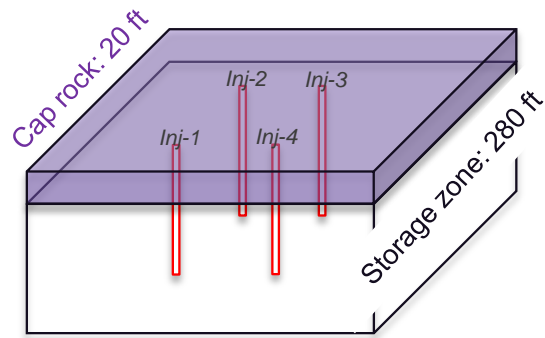

**Fig. 3. Schematic of the reservoir model for geological storage of $CO_2$.**

The schematic of the reservoir model is depicted in **Fig. 3**. There are 2 layers (layers 1 and 2) of "caprock" to seal $CO_2$ at the top, and 28 layers (layer 3 to 30) of "storage zone" to store $CO_2$. There are 4 $CO_2$ injection wells perforating all the 28 layers in the "storage zone", and in the $x - y$ plane they are drilled at grid indices (71, 71), (71, 141), (141, 141) and (141, 71), respectively. Two million tons of $CO_2$ was injected at a constant rate for 10 years. The permeability and porosity fields are correlated and assumed to be spatially heterogeneous and uncertain. 100 equiprobable realizations of permeability and porosity fields are generated following the same geological facies model. **Fig. 4** presents 5 examples of the permeability realizations of the storage zone and the corresponding porosity realizations. In each example the shapes of the high permeability and porosity zones are quite similar, but the magnitudes of permeability and porosity in these zones vary.

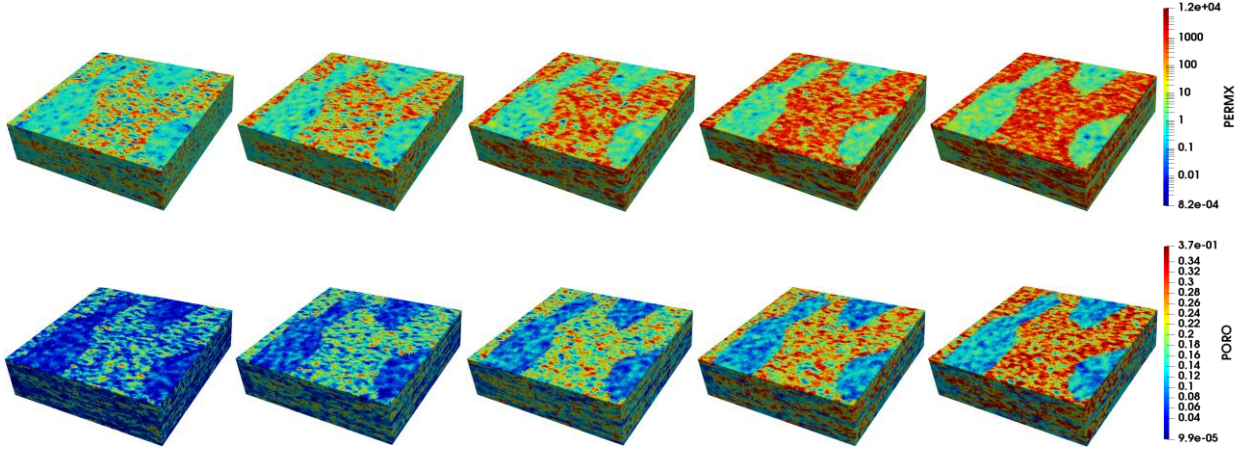

**Fig. 4. Realizations of permeability (upper row) and porosity (lower row) of the "storage zone". From left to right: example realizations of P25, P50, P75, P90 and P95, respectively.**

Simulations were performed on each of these 100 realizations with the commercial reservoir simulation, GEM from CMG (CMG, 2020) in order to generate training data for the DL workflow. The ensemble of realizations is split based on the permeability and porosity realizations where we choose 90 simulation runs for training, 5 runs for validation and 5 runs for testing.

### B. Training and Prediction Efficiency with Feature Coarsening

Since we have already demonstrated that feature coarsening can significantly reduce memory consumption in **Table 1**, we further investigate the impact of feature coarsening on training efficiency. Different strides were chosen to coarsen the feature images and feed them to train different FNO models. The training samples are divided into mini-batches with 20 samples/batch, and we train the FNO models with 100 epochs on a GPU (NVIDIA Quadro RTX 4000), and weight factor $\lambda$ in Equation (5) is set to 0.1 based on our sensitivity analysis.

**Table 2. Sensitivity of stride with training and prediction efficiency.**

| Stride | Feature Image Size | Training time, hours | Normalized Prediction Time*, sec |
|---|---|---|---|
| 1 | 211×211 | 40.23 | 0.023 |
| 2 | 106×106 | 10.47 | 0.24 |
| 3 | 73×73 | 9.02 | 0.12 |
| 4 | 55×55 | 8.78 | 0.076 |
| 5 | 43×43 | 8.86 | 0.066 |
| 10 | 22×22 | 8.36 | 0.041 |
| CMG's GEM | N/A | N/A | 168.75 |

\* Prediction time is normalized to predict a single snapshot (time step) of the 28 storage layers.

**Table 2** illustrates the training and prediction efficiency with different strides to coarsen the image size. The training at the full image size (211×211) takes 40.23 hours. As we coarsen the image by increasing the stride, the training time can be decreased by 74% (10.47 hours for stride = 2) to 79% (8.36 hours for stride = 10). This makes sense since smaller images require less computation in the backpropagation process for training FNO. Therefore, the feature coarsening makes the training process much more affordable as we are dealing with a large scale geological model. The prediction basically includes FNO prediction and resolution recovery by interpolation, except when stride is 1, where recovering the resolution is not necessary since FNO predicts at the original fine scale. Overall, the prediction time decreases as we coarsen the feature image, and the speedup compared to CMG's GEM simulation run is remarkable, at least 700 times (stride = 2).

**C. Prediction Accuracy**

To measure the temporal error of the DL workflow, we calculate the temporal error of pressure by,

$$p_{error}^t = \frac{1}{n_c n_g} \Sigma_{i=1}^{n_c} \Sigma_{j=1}^{n_g} \frac{\left\|p_{i,j}^t - \hat{p}_{i,j}^t\right\|}{\left(p_{i,j}^t\right)_{max} - \left(p_{i,j}^t\right)_{min}}, \quad (7)$$

where $t$ is the time step; $p$ is the pressure; $n_c$ is the number of testing simulation runs; and $n_g$ is the number of grid cells of the geological model.

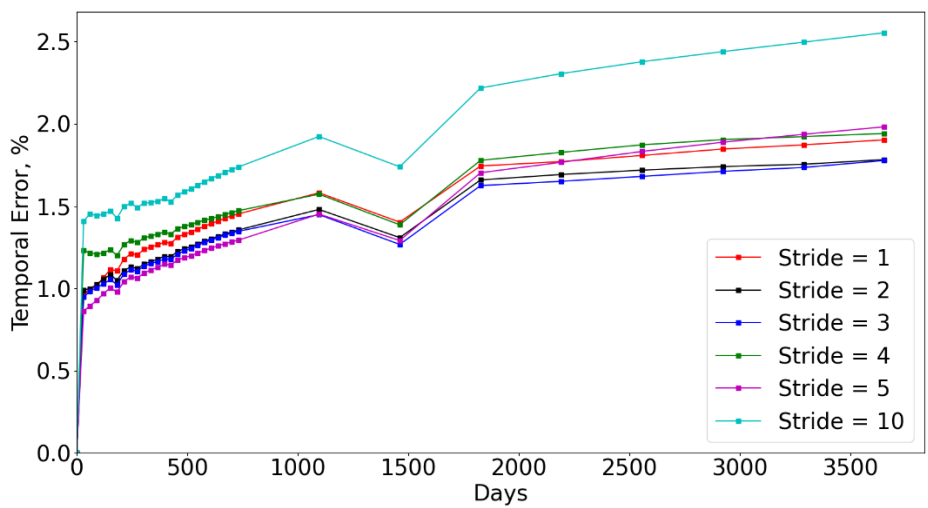

**Fig. 5. Temporal error propagation of pressure under different strides.**

In **Fig. 5**, we presents the temporal error of pressure at different strides. It shows that FNO (stride = 10) has the highest temporal error, which can be explained by fact that a larger stride bypasses more information in the feature images, and thus tends to lose more prediction fidelity. On the other hand, FNO (stride = 1) that predicts at the original fine scale does not actually lead to the highest accuracy (red line), which is likely due to the fact that the original fine-scale feature images, specifically permeability and porosity images, provide the FNO model with more information than necessary and make it not generalized as the scenarios with coarser resolution (e.g. stride = 2, 3). Ultimately FNO (stride = 3) (blue line) strikes the best balance between prediction accuracy (temporal error <1.5%) and training efficiency (9.02 hours), and makes predictions with a speedup of ~1400 compared to physics-based reservoir simulator.

Based on the pressure predicted by FNO (stride = 3), we calculate the mean relative error of each layer (from layer 3 to 30) by aggregating all the time steps in the 5 testing cases. **Fig. 6** shows that the mean relative errors remain low through the 28 storage layers, which gives confidence to use the DL workflow to predict the pressure evolution in 3D.

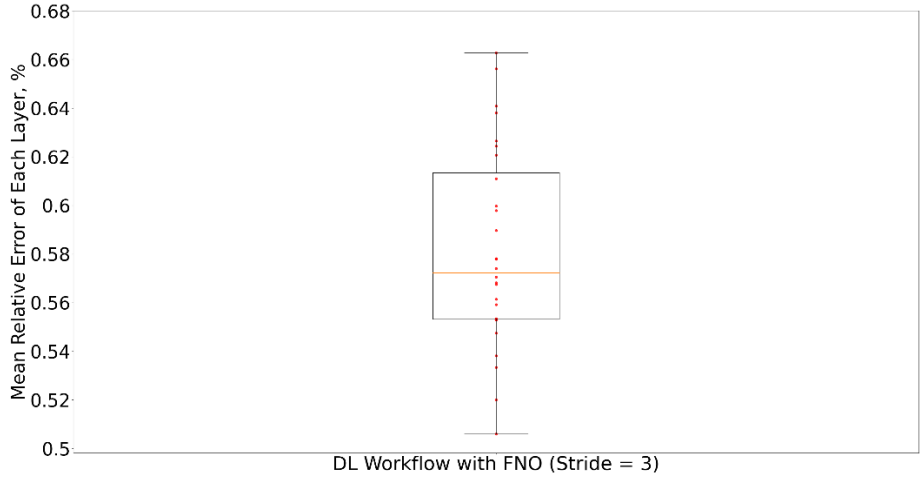

**Fig. 6. Performance of pressure prediction in 28 storage layers for the 5 testing cases.**

In **Fig. 7**, we present a representative example of the pressure fields at early and late times in the top (layer 3), middle (layer 16) and bottom (layer 30) layers in the "storage zone" for predictions of the DL workflow with FNO (stride = 3), and compare them with the ground truth from reservoir simulation. At the early period (1 year) the pressure plumes grow surrounding the 4 injection wells and the plume becomes larger with increasing depth (larger layer number) because of gravity. These details are captured by the predictions from our DL workflow (mean absolute error 2.826 psia and mean relative error 0.149%), and there are only small errors in the injection well locations. At the end of the $CO_2$ injection period (10 years) the pressure plume reaches its maximum size with the pressure error increasing slightly (mean absolute error 3.906 psia and mean relative error 0.202%). Besides, we observe that the DL workflow can generally predict smooth pressure field and delineates the irregular pressure plume shape driven by permeability and porosity heterogeneity quite well, especially for larger plume sizes.

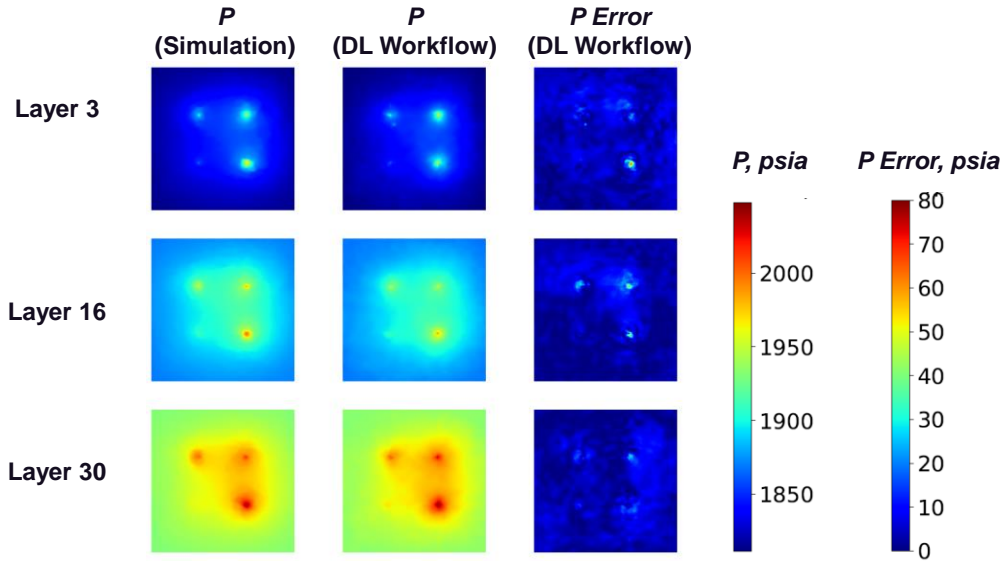

(a) Pressure after 1 year injection. Mean absolute error: 2.826 psia, mean relative error: 0.149%.

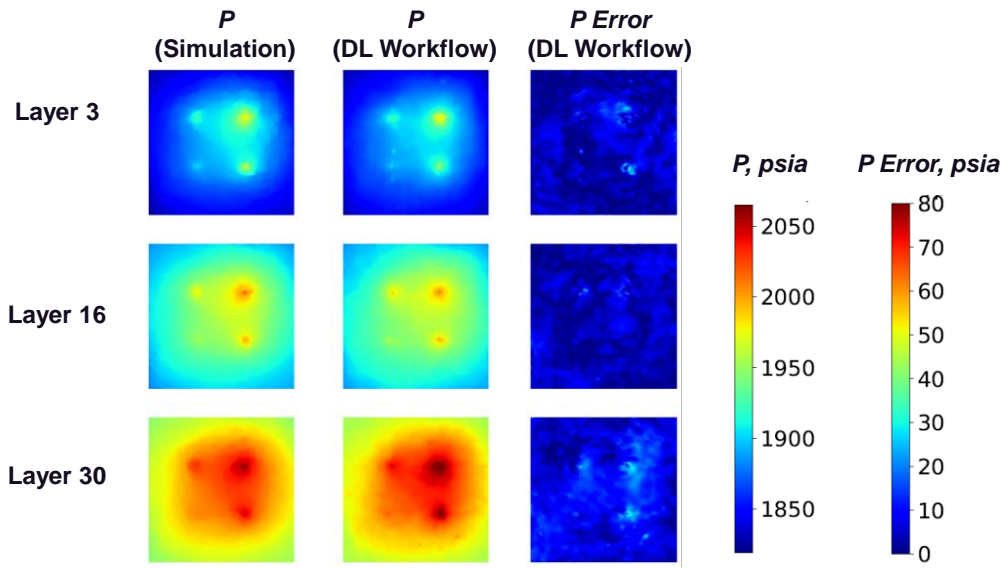

(b) Pressure after 10-year injection. Mean absolute error: 3.906 psia, mean relative error: 0.202%.

Fig. 7. Pressure prediction by the DL workflow with FNO (stride = 3).

## V. CONCLUSIONS

We developed a deep learning (DL) workflow to predict the pressure evolution as fluid flows in 3D large-scale heterogeneous porous media. With pre-defined stride, we performed feature coarsening to extract the most representative information of geology and well controls, which can help reduce memory consumption of feature arrays and improve training efficiency. Further, we recovered the resolution of the predicted pressure field to the fine scale based on the hypothesis of pressure continuity for fluid flow in porous media.

We evaluated the performance of the workflow by applying to the problem to a $CO_2$ injection into a large-scale 3D heterogeneous aquifer problem. We demonstrated that the feature coarsening procedure significantly reduces the memory consumption by more than 75% and decreases the training time by >74%, which can be reasonably interpreted as due to the fact that smaller feature image takes less computational time for the backpropagation in the training process. The feature coarsening process indeed causes some fidelity loss during the prediction. Nevertheless, the model trained at the original scale does not lead to the highest accuracy, which is likely to be caused by the fact that fine scale images provide the DL workflow with too detailed of information and lead to a loss of generalization. In 3D pressure prediction, we obtained decent temporal stability with temporal error <1.5%, and it is quite stable across all the layers. The DL workflow can delineate with pressure plume shape accurately with great smoothness, and even capture the plume size changes due to gravity. The speed of prediction by the DL workflow is ~1400 times faster than that of physics-based simulation, which is quite favorable for optimization and uncertainty quantification in many applications including $CO_2$ sequestration where physics-based simulations are computationally expensive.

## ACKNOWLEDGMENT

The authors acknowledge the financial support by US DOE's Fossil Energy Program Office through the project, Science-informed Machine Learning to Accelerate Real Time (SMART) Decisions in Subsurface Applications. Funding for SMART is managed by the National Energy Technology Laboratory (NETL).